# In-situ measurements of twist and bend elastic constants in oblique helicoidal cholesteric

Olena S. Iadlovska[12], Kamal Thapa[12], Mojtaba Rajabi[12], Sergij V. Shiyanovskii[13], and Oleg D. Lavrentovich[123*]

[1]*Advanced Materials and Liquid Crystal Institute, Kent State University, Kent, Ohio 44242, USA*
[2]*Department of Physics, Kent State University, Kent, Ohio 44242, USA*
[3]*Materials Science Graduate Program, Kent State University, Kent, Ohio 44242, USA*
*e-mail address: olavrent@kent.edu

**Abstract**

Unique electro-optical properties of the oblique helicoidal cholesteric (Ch$_{OH}$) stem from its heliconical director structure. An applied electric field preserves the single harmonic modulation of the director while tuning the Ch$_{OH}$ period and the corresponding Bragg peak wavelength within a broad spectral range. We use the response of Ch$_{OH}$ to the electric field to measure the elastic constants of twist $K_{22}$ and bend $K_{33}$ directly in the cholesteric phase. The temperature dependencies of $K_{22}$ and $K_{33}$ allow us to determine the range of the electric tunability of the Ch$_{OH}$ pitch and the heliconical angle. The data are important for understanding how molecular composition and chirality influence macroscopic elastic properties of the chiral liquid crystals and for the development of Ch$_{OH}$-based optical devices.

## INTRODUCTION

Elastic properties of liquid crystals are determined from the materials' response to external forces such as mechanical stress, electric and magnetic fields [1,2]. In the nematic (N) liquid crystal, the elastic constants are typically measured by using a uniformly aligned material and applying an electric or magnetic field to create deformations of splay, twist, or bend [3,4]. Application of the same methods to the chiral analog of N, called a cholesteric (Ch), is difficult since the ground state is not uniform: the director twists in space, being perpendicular to the helicoidal axis. A notable exception is the measurement of the Ch twist elastic constant $K_{22}$ by a complete unwinding of the helicoidal structure by an electric or magnetic field [5]. It is often assumed that the elastic constants of Ch have the same values as the ones measured in the N analogs. Such an approximation worsens when chiral molecular interactions become stronger.

Recent research [6] demonstrated that the bend elastic constant $K_{33}$ of Ch could be determined *in situ* if the material forms the so-called oblique helicoidal cholesteric (Ch$_{OH}$) structure when exposed to external electromagnetic fields [6-12]. In Ch$_{OH}$, the director twists around a single axis that is parallel to the applied field, but unlike in a conventional Ch, it is also tilted with respect to the helicoidal axis, Fig.1. The structure contains both twist and bend of the director. The Ch$_{OH}$ state, predicted theoretically [7,8] and confirmed experimentally [9-11], forms in materials with a low $K_{33}$, such as dimers with flexible aliphatic spacers [13-17]. The period $P$ and the conical tilt angle $\theta$ of the Ch$_{OH}$ structure are defined by the strength of the field and by the material parameters such as the dielectric anisotropy, intrinsic pitch, and two elastic constants, $K_{22}$ and $K_{33}$.

In this work, we demonstrate that the unique structure of Ch$_{OH}$ allows one to measure both $K_{22}$ and $K_{33}$ and their temperature dependencies in the Ch material. The measurement of $K_{33}$ requires one to determine the dielectric anisotropy, refractive indices, and the dependence of the Bragg reflection wavelength on the applied electric field under the normal incidence of light. The $K_{22}$ measurements are more involved and require one to determine the intrinsic (field-free) pitch $P_0$, the critical electric field $E_{NC}$, at which Ch$_{OH}$ unwinds into a uniform structure, and $K_{33}$. $E_{NC}$ is determined by two different approaches, based on capacitance and interference measurements, which allow one to detect a very small Ch$_{OH}$ pitch (below 50 nm) at the fields approaching $E_{NC}$. The new $E_{NC}$ measurement techniques are advantageous as compared to the Bragg reflection or microscopy observations of textures, which are adapted to much larger values of the pitch [9]. Once the moduli $K_{22}$ and $K_{33}$ are measured, other properties could be deduced, such as the range of field tunability of the Ch$_{OH}$ pitch $P$ and the conical angle $\theta$. The proposed protocol thus yields all the static material properties of Ch$_{OH}$, except for the splay elastic constant.

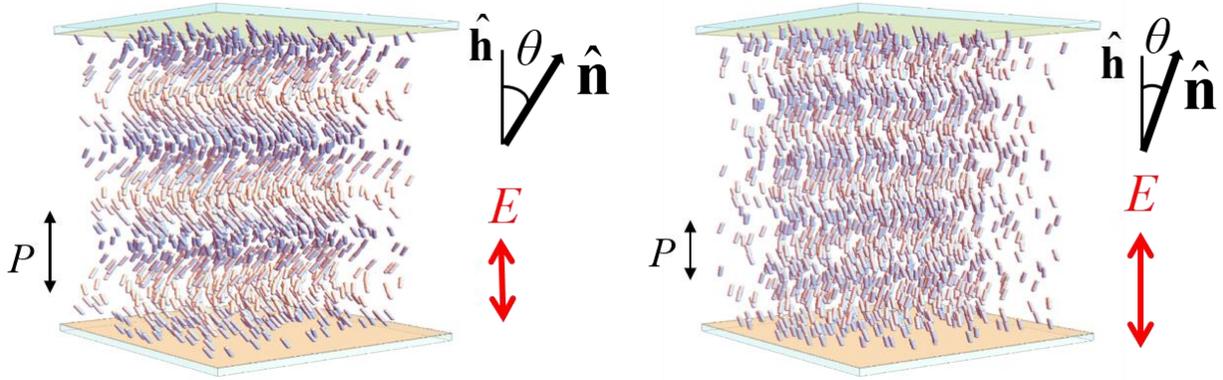

FIG. 1. The oblique helicoidal cholesteric structure Ch$_{OH}$ in an applied electric field: the Ch$_{OH}$ pitch $P$ and the conical angle $\theta$ both decrease when the electric field **E** increases.

## MATERIALS AND EXPERIMENTAL TECHNIQUES

*N and Ch materials*. A room-temperature N mixture is formulated using flexible dimers 1″,7″-bis(4-cyanobiphenyl-4′-yl) heptane (CB7CB) and 1-(4-cyanobiphenyl-4′-yloxy)-6-(4-cyanobiphenyl-4′-yl) hexane (CB6OCB) (both purchased from SYNTHON Chemicals GmbH & Co. KG), and a rod-like mesogen pentyl cyanobiphenyl (5CB) (EM Industries), in the weight proportion 5CB:CB7CB:CB6OCB = 52:31:17. The dimer molecules show bend conformations which yields a small $K_{33}$ [13-17], while 5CB shifts the temperature range of the mesophase down to room temperature. The N mixture is doped with the left-handed chiral agent S811 (EM Industries), 5CB:CB7CB:CB6OCB:S811 = 50:30:16:4, to produce the Ch mixture. The mixture without the chiral dopant shows the phase diagram N$_{TB}$ (17.4 °C) N (78 °C) Iso, which is noticeably different from the phase diagram of the Ch mixture: N$_{TB}$* (15.5 °C) Ch (66.3 °C) Iso. Here, N$_{TB}$ is the so-called twist-bend nematic phase and N$_{TB}$* is its chiral analog, Iso stands for the isotropic phase. The difference suggests that the material properties of the Ch with chiral additive are not the same as those of the achiral N base. The phase diagrams are measured on cooling using Linkam hot stage equipped with the PE94 temperature controller (both Linkam Scientific) and EHEIM Professional-3 cooling system (EHEIM GmbH & Co. KG).

*Cells with preset surface alignment*. Three types of flat sandwich cells were studied, with the planar, homeotropic and hybrid anchoring [18] of the local director at the bounding surfaces. Planar cells with indium tin oxide (ITO) electrodes and gap thicknesses $d = (4 – 23)$ μm were either purchased from E.H.C. Co, Japan, or assembled in the laboratory. Planar alignment is achieved by a rubbed layer of polyimide PI2555 (Nissan Chemicals, Ltd.). Homeotropic cells, $d = (2 - 4)$ µm, are prepared in the laboratory using the technique developed by Young-Ki Kim [19,20]. The reactive mesogen RM257 is mixed with the homeotropic aligning agent SE5661 and doped with the UV photo-initiator Irgacure 651, in weight proportion SE5661:RM257:Irgacure 651 = 96.7:3:0.3. The mixture is spin-coated onto an ITO glass substrate and baked at 170 °C for 1 hour. The substrates are then irradiated with UV light (365 nm) for 90 minutes [19,20]. The cell thickness is set by spherical spacers mixed with UV-curable adhesive NOA68 (Norland Products, Inc.), and the thickness is measured using Lambda 18 UV/VIS spectrometer (Perkin Elmer, Inc). The cells are filled with N and Ch mixtures in the isotropic phase and slowly cooled down (0.2°C/min) to the desired temperature.

*Electric field control of the structure.* To create the $Ch_{OH}$ state, an alternate current (ac) field with a frequency 3 kHz is applied to the ITO electrodes using Keithley 3390 waveform function generator (Keithley Instruments) and wideband Krohn-Hite 7602M amplifier (Krohn-Hite Co.). In the applied field, the helicoidal axis $\hat{\mathbf{h}}$ is parallel to the field $\mathbf{E}$, $\hat{\mathbf{h}} \parallel \mathbf{E}$, and the $Ch_{OH}$ pitch $P$ is tunable by the field in a broad range [7,8].

The electric field does not cause noticeable heating of cells. A thin homeotropic N cell, $d = 4$ µm, was subject to a 3 kHz ac field, ranging from 1 V/µm to 5 V/µm. Phase transition temperatures measured at different applied voltages showed no changes exceeding 0.1°C. Other material parameters such as the parallel component $\varepsilon_\parallel$ of dielectric permittivity, did not change by more than 2%. This 2% difference is equivalent to a temperature shift by less than 0.05°C.

*Spectral measurements: oblique and normal incidence of light.* Optical spectra are measured using a tungsten halogen light source LS-1 with a working range 350-2000 nm and a USB2000 Vis-NIR spectrometer (both Ocean Optics, Inc.). At the normal incidence of light, unpolarized light from the light source is focused by a lens into a paraxial ray. The reflected beam is passed back through the bifurcated fiber and detected by the spectrometer interfaced with the OceanView spectroscopy software. The Bragg reflection wavelength is measured as a function of the applied electric field, as described previously [6].

In addition to the normal incidence measurements, we use the Bragg reflection for the oblique incidence of light to verify the values of the intrinsic pitch $P_0$ and the critical electric field $E_{NC}$ (at which the helix is unwound) measured by other techniques (see below). At the oblique incidence of light, Bragg reflection spectra are measured using a fiber optics setup based on a rotating goniometer stage (Euromex Holland), Fig 2(a). Polarization of the incident and reflected beams is set by a pair of linear wire-grid polarizers (Thorlabs, Inc.) placed along the optical path at the exit from the fiber connected to the light source and at the entrance to the fiber connected to the USB2000 detector. p- and s-polarization states of an electromagnetic wave are defined as the electric field components, $\mathbf{E}_p$ and $\mathbf{E}_s$, parallel and perpendicular to the plane of incidence, respectively. The Ch$_{OH}$ cell is mounted in the center of the goniometer stage such that the plane of incidence is parallel to the horizontal $xz$-plane. The angle of incidence $\beta$, measured from the normal $\hat{\mathbf{z}}$ to the Ch$_{OH}$ cell, is set with an accuracy better than 0.25°. The Ch$_{OH}$ cell is sandwiched between two semi-cylindrical prisms (N-BK7 glass, $n_p$ = 1.518 at 547 nm, DelMar Photonics). A Cargille matching liquid with the refractive index $n$ = 1.516 (at 600 nm) fills the gaps between the cell substrates and prisms to reduce reflection losses. The semi-cylindrical prisms allow one to measure light reflection in a broad angular range and assure that the angle of incidence from the glass is the same as the incidence angle from the air: $\beta_g = \beta_{air} \equiv \beta$.

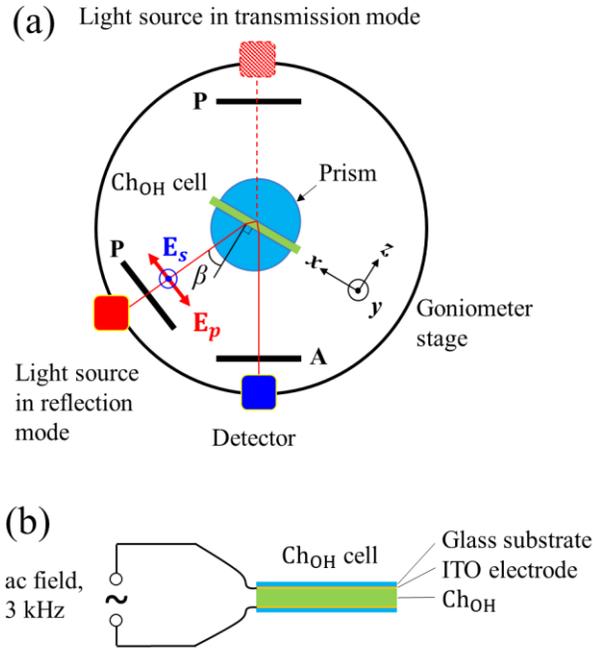

FIG. 2. (a) Reflection (solid lines) and transmission (dashed lines) modes of the oblique incidence setup (top view). The $xz$ plane of incidence is parallel to the goniometer stage, and the angle of incidence $\beta$ is measured from the normal to the Ch$_{OH}$ cell. The polarizations of the incident and reflected rays are set by the polarizer (P) and analyzer (A), respectively. (b) Ch$_{OH}$ cell geometry.

Besides the wavelength of Bragg reflection, the determination of $K_{22}$ and $K_{33}$ also requires the knowledge of (1) dielectric anisotropy $\Delta\varepsilon$, (2) ordinary $n_o$ and extraordinary $n_e$ refractive indices, (3) $P_0$, and (4) $E_{NC}$. Below we present the corresponding techniques.

*Dielectric anisotropy* $\Delta\varepsilon$. The dielectric characterization of Ch is performed using an LCR meter 4284A (Hewlett Packard). The parallel permittivity $\varepsilon_\parallel$ is determined from the capacitance of a homeotropic cell, $d = 1.9$ µm, measured at a high electric field $E > E_{NC}$, and the perpendicular permittivity $\varepsilon_\perp$ is measured in the planar cell, $d = 9.1$ µm, by applying a weak electric field $E = 10^{-3}$ V/µm that does not perturb the planar Ch structure. The dielectric anisotropy $\Delta\varepsilon$ is then calculated as the difference $\varepsilon_\parallel - \varepsilon_\perp$, Fig.3.

*Birefringence and refractive indices.* Birefringence is measured in both the Ch and N mixtures. Ch birefringence is calculated as $\Delta n = \Gamma/d$, Fig.4(b), where $\Gamma$ is the optical retardance of a thin ($d = 5.3$ µm) planar cell with the director unwound by a strong in-plane ac (3 kHz) electric field $E > E_{NC}$, applied to two indium tin oxide (ITO) stripe electrodes separated by a 100 µm gap, Fig.4(a). $\Gamma$ is measured using the Exicor MicroImager System (Hinds Instruments) at three different wavelengths 475, 535, and 655 nm.

In the N mixture, the ordinary $n_o$ and extraordinary $n_e$ refractive indices are measured using a wedge cell and the interference technique [21]. The wedge cell is assembled from two glass substrates with ITO electrodes coated with rubbed polyimide PI2555 (Nissan Chemicals, Ltd.) to achieve planar alignment. The rubbing direction is perpendicular to the thickness gradient. The thickness of the thick end of the cell is set by 100 µm spacers mixed with the adhesive NOA68. There are no spacers at the glued wedge. The wedge angle is less than 1°. The cell is filled with the N mixture in the isotropic phase, and the refractive indices are measured at 532 nm on cooling. The $\Delta n$ values of the Ch and N phases are in very good agreement with each other, Fig.4(b). Therefore, we used the N mixture to determine the dispersion of the ordinary refractive index, which is needed for the extraction of $K_{33}$ from the Bragg spectra.

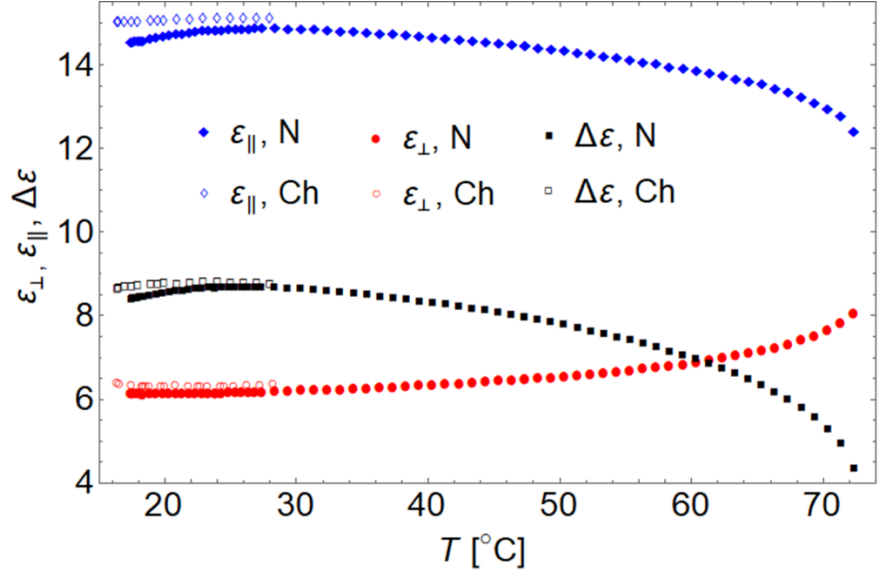

FIG. 3. Temperature dependencies of dielectric permittivities $\varepsilon_\parallel$, $\varepsilon_\perp$ and $\Delta\varepsilon$ measured in the N and Ch mixtures (filled and empty symbols, respectively).

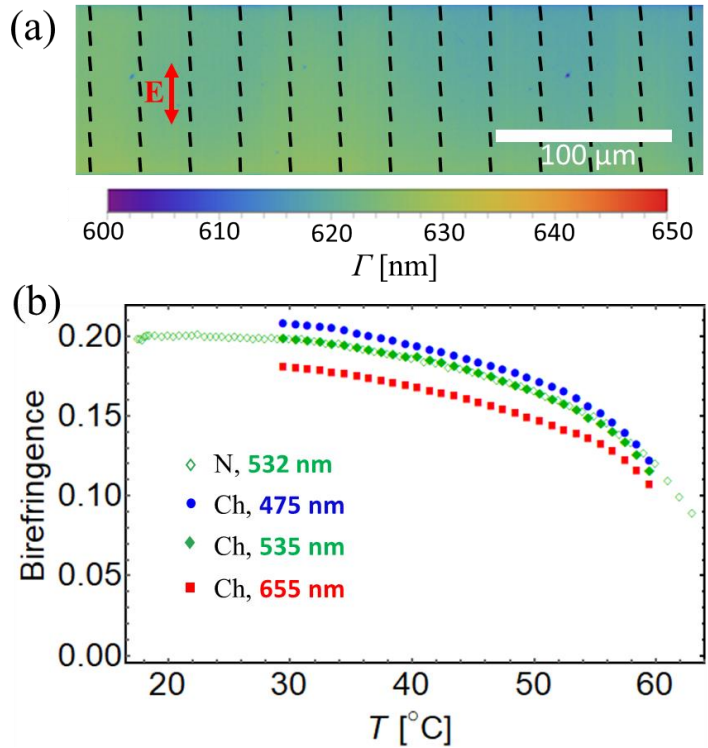

FIG. 4. (a) MicroImager map of the optical retardance of the Ch$_{OH}$ cell subject to a high in-plane electric field $E > E_{NC}$. (b) Temperature dependence of the birefringence measured in N (empty symbols) and Ch (filled symbols) mixtures. The size of all data symbols exceeds the error of measurement.

The dispersion $n_o(\lambda)$ is described by the Cauchy formula $n_o(\lambda) = A + B\lambda^{-2} + C\lambda^{-4}$. We measure the temperature dependence of $n_o$ at 488, 532, and 633 nm, using laser line color filters with 1 nm central bandwidth (Thorlabs, Inc.). Within the temperature range $3.5°C \leq T - T_{TB} \leq 6°C$, $n_o$ is practically temperature-independent, with the mean values $n_o = 1.564$ at 488 nm, $n_o = 1.560$ at 532 nm, and $n_o = 1.552$ at 633 nm. The polynomial coefficients found by fitting are $A = 1.603$, $B = -2.12 \cdot 10^{-3}$ µm$^2$, and $C = 2.15 \cdot 10^{-3}$ µm$^4$.

*Cholesteric pitch $P_0$*. The temperature dependence of the intrinsic Ch pitch $P_0$ is measured in the absence of the electric field using the Grandjean-Cano wedge cell. The cell is assembled from two ITO glass substrates coated with a rubbed layer of polyimide PI2555 to achieve planar alignment. The rubbing direction is perpendicular to the thickness gradient. The thickness of the thick end of the cell is set by 140 µm glass slides mixed with the adhesive NOA68, and there are no spacers at the glued wedge. The wedge angle is 0.83°. $P_0$ is determined from the spacing of dislocations of a Burgers vector $P_0/2$ [22]; it increases as the temperature is lowered, Fig. 5 (a).

Since in some cases the Grandjean-Cano wedge approach could produce incorrect results, if attention is not being paid to the Burgers vector of the dislocations [22] and to their locations [23,24], we verified the value of $P_0$ in an independent experiment, by measuring the wavelength of Bragg reflection peaks at oblique incidence for a planar Ch cell with $d = 16$ µm at $T = 19$ °C, Fig. 5 (b). At the fixed angle of incidence $\beta = 65°$, we use the p-polarized incident beam and s-polarized reflected beam to obtain the spectrum with two reflection maxima $\lambda = \bar{n} \left(\frac{P_0}{m}\right) \cos \beta_{LC}$ that correspond to $m = 4$ and 6, i.e., to $P_0/4$ and $P_0/6$ values; the angle of light propagation in the liquid crystal $\beta_{LC}$ is calculated using Snell's law, $\beta_{LC} = arcsin[(n_g/\bar{n}) \sin \beta]$, where $\bar{n}$ is the average of refractive indices, and $n_g = 1.52$ (at 600 nm) is the refractive index of the soda-lime glass. $P_0$ deduced from the spectra coincides with the data obtained from the Grandjean-Cano wedge, Fig.5 (a).

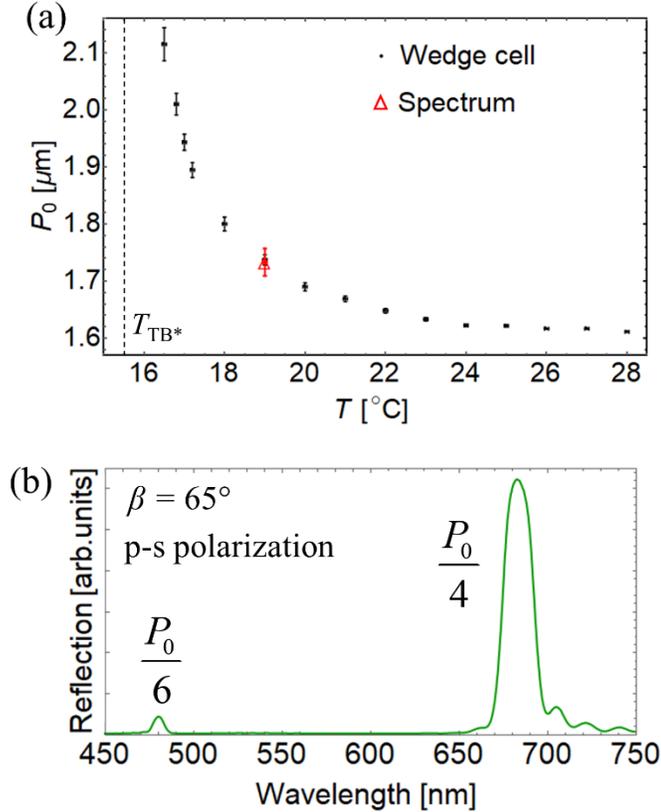

FIG. 5. (a) Temperature dependence of the pitch $P_0$ ($T$) measured in the wedge cell and (b) Bragg reflections corresponding to $P_0/4$ and $P_0/6$; oblique incidence of light, Ch$_{OH}$ cell thickness $d = 16$ µm.

*Critical field $E_{NC}$ of unwinding the Ch$_{OH}$ structure.* The critical field $E_{NC}$, at which the helix is unwound, is determined by measuring the capacitance of thin homeotropic Ch$_{OH}$ cells, $d = 1.9$, 3.3, and 3.9 µm while changing the applied electric field, Fig.6 (a). At $E = E_{NC}$, the conical angle vanishes and does not change if the field increases further, which results in a clearly visible kink. The change of capacitance with the field at $E > E_{NC}$ is caused by the suppression of director fluctuations [25].

The critical field $E_{NC}$ was also determined by measuring the transmission spectra through a planar Ch$_{OH}$ cell, $d = 22.7$ µm, for an oblique incidence, $\beta = 50°$. The polarizer P and analyzer A are either parallel or crossed, and both are oriented at 45° with respect to the $xz$ plane of incidence. The field dependency of the light interference extremum shows a similar "kink" at $E = E_{NC}$, Fig.6(b). Both methods yield very close temperature dependencies of $E_{NC}$, Fig.6(c). The agreement provides an additional confirmation that the electric field does not cause noticeable heating of cells. Note here that in the studied mixture, the pitch stabilized by the electric fields approaching $E_{NC}$ is very small, decreasing below 50 nm, Fig.10 (b); therefore, the capacitance and interference methods are better suited for the measurements of $E_{NC}$ as compared to Bragg reflection or polarizing microscopy observations of the fingerprint textures [9].

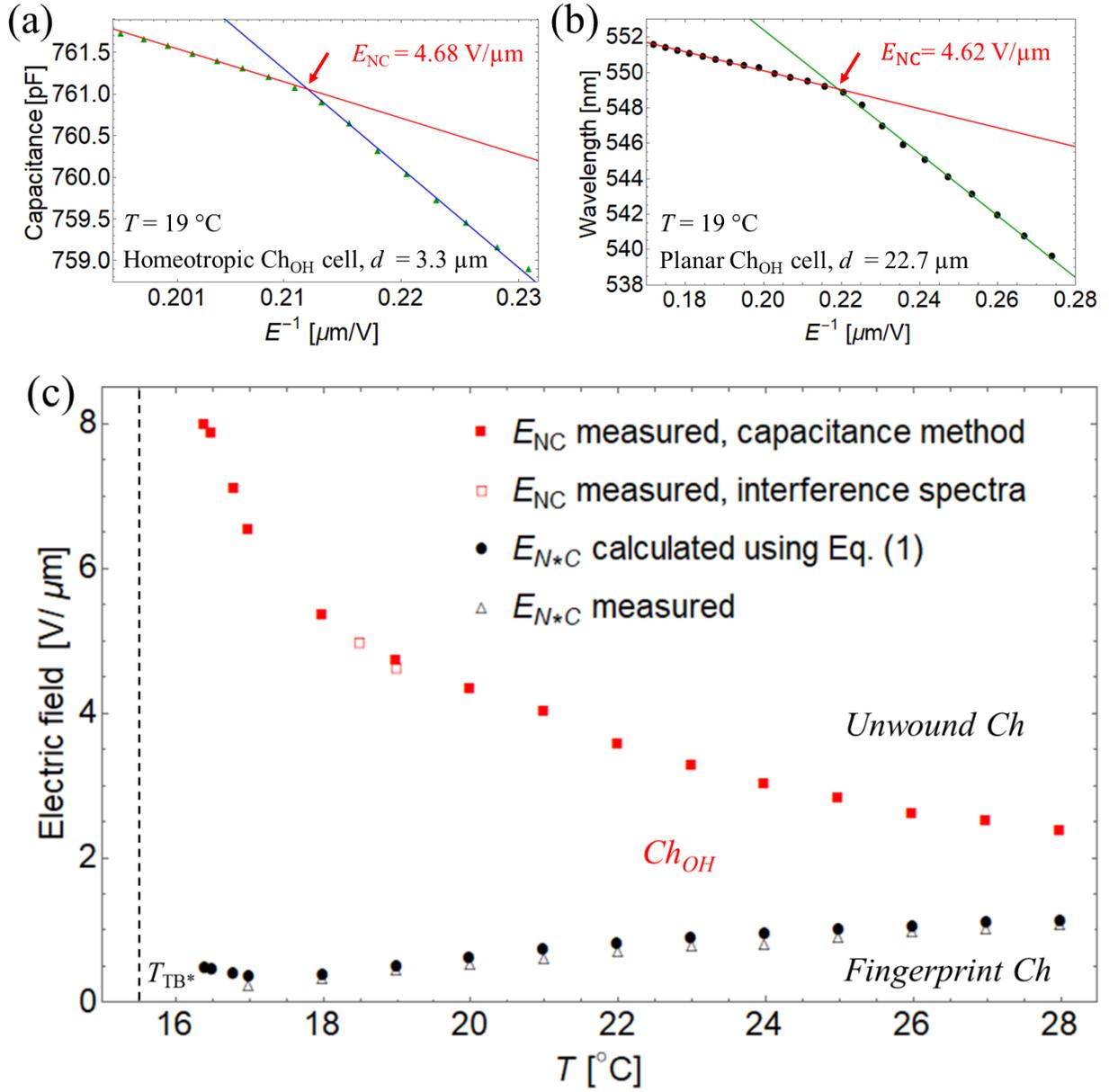

FIG. 6. (a) Capacitance and (b) interference maxima vs electric field; homeotropic Ch$_{OH}$ cell, $d$ = 3.3 µm, in the capacitance measurements and the planar Ch$_{OH}$ cell, $d$ = 22.7 µm, in the spectral measurements. The "kinks" correspond to the critical field $E_{NC}$. (c) Temperature dependence of the critical fields $E_{NC}$ and $E_{N*C}$; $E_{N*C}$ is measured experimentally and deduced from Eq. (1).

*Critical field $E_{N*C}$ of the $Ch_{OH}$-to-Ch transition.* Depending on the applied electric field, the Ch cells show three states: an unwound state at $E > E_{NC}$, a $Ch_{OH}$ state at $E_{N*C} < E < E_{NC}$, Fig.7 (a), and a Ch state with the director perpendicular to the helicoidal axis, at $E < E_{N*C}$ [9], Fig.7 (b,c). During the first-order $Ch_{OH}$-to-Ch structural transition in a hybrid aligned cell, $d = 20.1$ µm, caused by the lowering of the electric field, the helical axis $\hat{\mathbf{h}}$ realigns from being parallel to $\mathbf{E}$ in the $Ch_{OH}$ to being perpendicular to it, $\hat{\mathbf{h}} \perp \mathbf{E}$, thus forming a polydomain Ch texture, Fig. 7 (c).

The polydomain Ch texture in Fig.7 (c) resembles polydomain textures (often called focal conic domain textures) in Ch cells with homeotropic anchoring at both plates. The similarity is natural since a Ch material with a positive dielectric anisotropy tends to align its helicoidal axis perpendicular to the electric field, in contrast to the $Ch_{OH}$ state, in which the helicoidal axis is aligned along the electric field. Furthermore, in the absence of the electric field, the director and thus the Ch helicoidal axis are subject to conflicting homeotropic and planar anchoring conditions at the opposite plates. In the $Ch_{OH}$ state, at $E_{N*C} < E < E_{NC}$, it is the external electric field that aligns the helicoidal axis and stabilizes the heliconical director with twist and bend; the conflicting surface anchoring at the plates results in minor changes of the director field, making the conical angle $\theta$ somewhat smaller than its equilibrium bulk value near the homeotropic plate and larger near the planar plate; the thickness of the subsurface layers where $\theta$ is different from the bulk value is less than 1 µm [26]. As demonstrated in the previous study [26], this subsurface change could affect the equilibrium field-stabilized pitch, which could decrease by up to ~10 nm in a homeotropic cell and increase by up to ~10 nm in a planar cell. To account for these effects in the measurement of $E_{N*C}$, we use a hybrid aligned cell, in which the opposite trends mitigate each other, and the overall pitch is close to its equilibrium bulk value expected in infinitely thick samples.

The measured $E_{N*C}$, Fig. 6(c) is in good agreement with the theoretical prediction [9],

$$E_{N*C} = E_{NC} \frac{\kappa\left[2+\sqrt{2(1-\kappa)}\right]}{1+\kappa}, \tag{1}$$

based on the comparison of the free energies of the $Ch_{OH}$ and Ch structures; in the evaluation of Eq. (1), we use the elastic ratio $\kappa = K_{33}/K_{22}$ that is presented in the next section.

Both $E_{NC}$ and $E_{N*C}$ define the electric field limits of the Ch$_{OH}$ state. The range of the Ch$_{OH}$ stability widens when one lowers the temperature towards the N*$_{TB}$ phase.

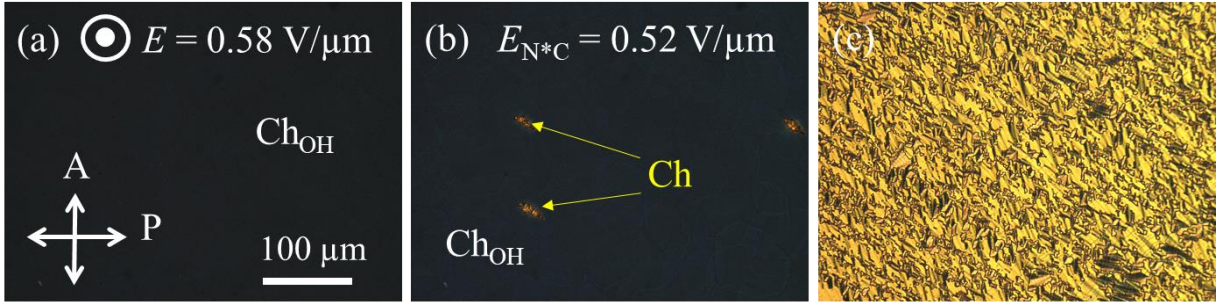

FIG. 7. Polarizing optical microscopy of the Ch$_{OH}$ to Ch transition at $T = 20$ °C; (a) Ch$_{OH}$ texture at $E = 0.58$ V/μm; (b) nucleation of Ch texture, 30 sec after the field is reduced to $E_{N*C} = 0.52$ V/μm; (c) expansion of the Ch texture at the same temperature and the applied field, 8 min after the start of nucleation.

## ELASTIC CONSTANTS OF BEND $K_{33}$ AND TWIST $K_{22}$

The bend constant $K_{33}$ is deduced from the Bragg reflection spectra when unpolarized light is incident normally at a planar Ch$_{OH}$ cell, as described in Ref. [6]. The Bragg wavelength $\lambda_{Bragg}$ is related to the Ch$_{OH}$ pitch $P$ and $K_{33}$ as [9]:

$$\lambda_{Bragg} = \bar{n}_{eff} P, \qquad \text{and } P = \frac{2\pi}{E}\sqrt{\frac{K_{33}}{\varepsilon_0 \Delta\varepsilon}}, \qquad (2)$$

where $\bar{n}_{eff} = (n_o + n_{e,eff})/2$ and $n_{e,eff} = n_o n_e / \sqrt{n_e^2 \cos^2\theta + n_o^2 \sin^2\theta}$; for small $\theta$, $n_{e,eff} \approx n_o \left(1 + \frac{1}{2}\left(1 - \frac{n_o^2}{n_e^2}\right)\sin^2\theta\right)$. The experimental points $\lambda_{Bragg}/n_o(\lambda_{Bragg})$ are fitted with the polynomial,

$$\frac{\lambda_{Bragg}}{n_o(\lambda_{Bragg})} = a_1 E^{-1} + a_2 E^{-2} + O(E^{-3}); \qquad (3)$$

here $E = U/d$, $U$ is the applied voltage and $n_o$ is interpolated to $\lambda_{Bragg}$ using the experimental data in Fig.4 (b). As demonstrated in Ref. [6], the fitting parameters are such that the $a_2$ term in Eq.(3) could be neglected, $a_1/a_2 E \sim 10^2$, and $K_{33}$ is calculated as

$$K_{33} = \frac{\varepsilon_0 \Delta \varepsilon}{4\pi^2} a_1^2. \quad (4)$$

Figure 8 shows a typical fit of the experimental data with Eq. (3), which confirms that in our experiments, the $a_2$ term does not affect the deduced $K_{33}$. Figure 9(a) shows the temperature dependency of $K_{33}$; the accuracy of the measured $K_{33}$ is better than 3% in both Figs.8 and 9(a). To explore whether the boundary conditions could affect the accuracy, the spectra were analyzed also for hybrid cells in the temperature range 16.5 – 22.0 °C. The two sets of $K_{33}$ data differ less than by 2%.

The twist elastic constant $K_{22}$ is found from the expression [9]

$$K_{22} = \frac{P_0 E_{NC} \sqrt{\varepsilon_0 \Delta \varepsilon K_{33}}}{2\pi}; \quad (5)$$

its temperature dependence is shown in Fig.9(a). Since both the intrinsic pitch $P_0$ and $K_{22}$ are known, we could deduce the temperature dependence of the chiral parameter $K_C = 2\pi K_{22}/P_0$, which could also be expressed as $K_C = E_{NC}\sqrt{\varepsilon_0 \Delta \varepsilon K_{33}}$ using Eq. (5), see Fig.9(b).

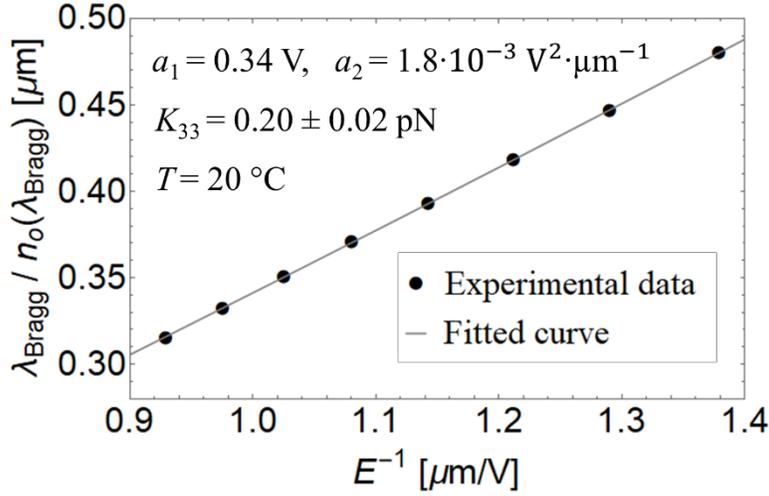

FIG.8. The field dependence of Bragg reflection wavelength used to calculate $K_{33}$.

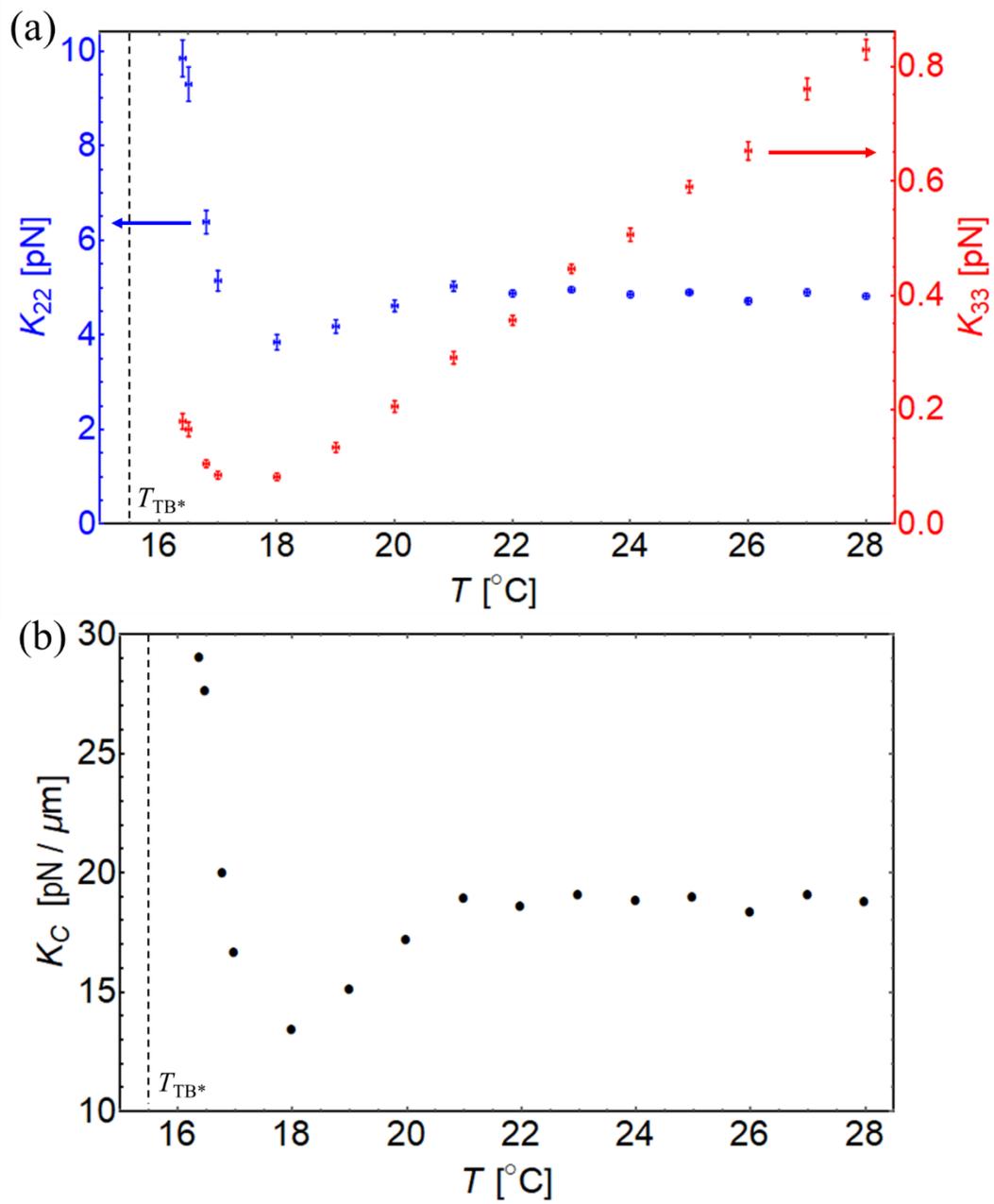

FIG. 9 (a) Temperature dependencies of $K_{33}$ and $K_{22}$, and (b) chiral parameter $K_C$.

The Ch$_{OH}$ structure depends critically on $K_{22}$ and $K_{33}$. The effect is understandable since Ch$_{OH}$ could exist only if $K_{33}$ is sufficiently small to allow the heliconical structure with both twist and bend. The elastic anisotropy ratio $\kappa = K_{33}/K_{22}$ determines the range of the electric field tunability of the Ch$_{OH}$ pitch $P$ and the conical angle $\theta$ [9],

$$P = \frac{\kappa E_{NC} P_0}{E}, \quad (6)$$

$$\sin^2\theta = \frac{\kappa}{1-\kappa}\left(\frac{E_{NC}}{E} - 1\right). \quad (7)$$

The elastic ratios $\kappa$ and $\kappa/(1-\kappa)$ in Eqs. (6)-(7) show a strong temperature dependency, Fig.10(a), which implies that the optical properties such as the Bragg reflection wavelength (related to $P$) and the width of the Bragg peak (related to $\theta$) are also temperature-dependent. Substituting $E = E_{N*C}$ and $E = E_{NC}$ into Eq. (6), one can determine the largest $P_{N*C} = \frac{(1+\kappa)P_0}{2+\sqrt{2-2\kappa}}$ and the smallest $P_{NC} = \kappa P_0$ pitch, Fig. 10(b). The tunable Ch$_{OH}$ pitch range $P_{N*C} < P < P_{NC}$ is very broad near the Ch-N$_{TB*}$ transition temperature, where the ratio $P_{N*C}/P_{NC}$ is about 18, Fig.10(c), which means that the electric field could tune the wavelength of Bragg reflection in an extraordinary broad range, from some $\lambda_{min}$ to $\lambda_{max} \approx 18\,\lambda_{min}$. Broadening of the Ch$_{OH}$ pitch range is caused by a steep decrease of $K_{33}$ near the Ch-N$_{TB*}$ phase transition. Finally, the smallest field at which the Ch$_{OH}$ state still exists, $E = E_{N*C}$ in Eq. (7), defines the widest cone angle of the heliconical structure $\theta_{N*C}$, which reaches about 32.4° near the transition to the N$_{TB*}$ phase, Fig. 10(d). Since our experiments confirm the validity of Eq.(1) for $E_{N*C}$, see Fig.6 (c), we could use Eq.(1) and Eq.(6) to demonstrate that $\theta_{N*C}$ is fully determined by the elastic ratio $\kappa$:

$$\cos^2\theta_{N*C} = \frac{1}{\sqrt{2(1-\kappa)}}. \quad (8)$$

In the limit $\kappa = 0$, the widest possible $\theta_{N*C}$ is 32.8°; if $\kappa > 1/2$, the Ch$_{OH}$ state does not occur.

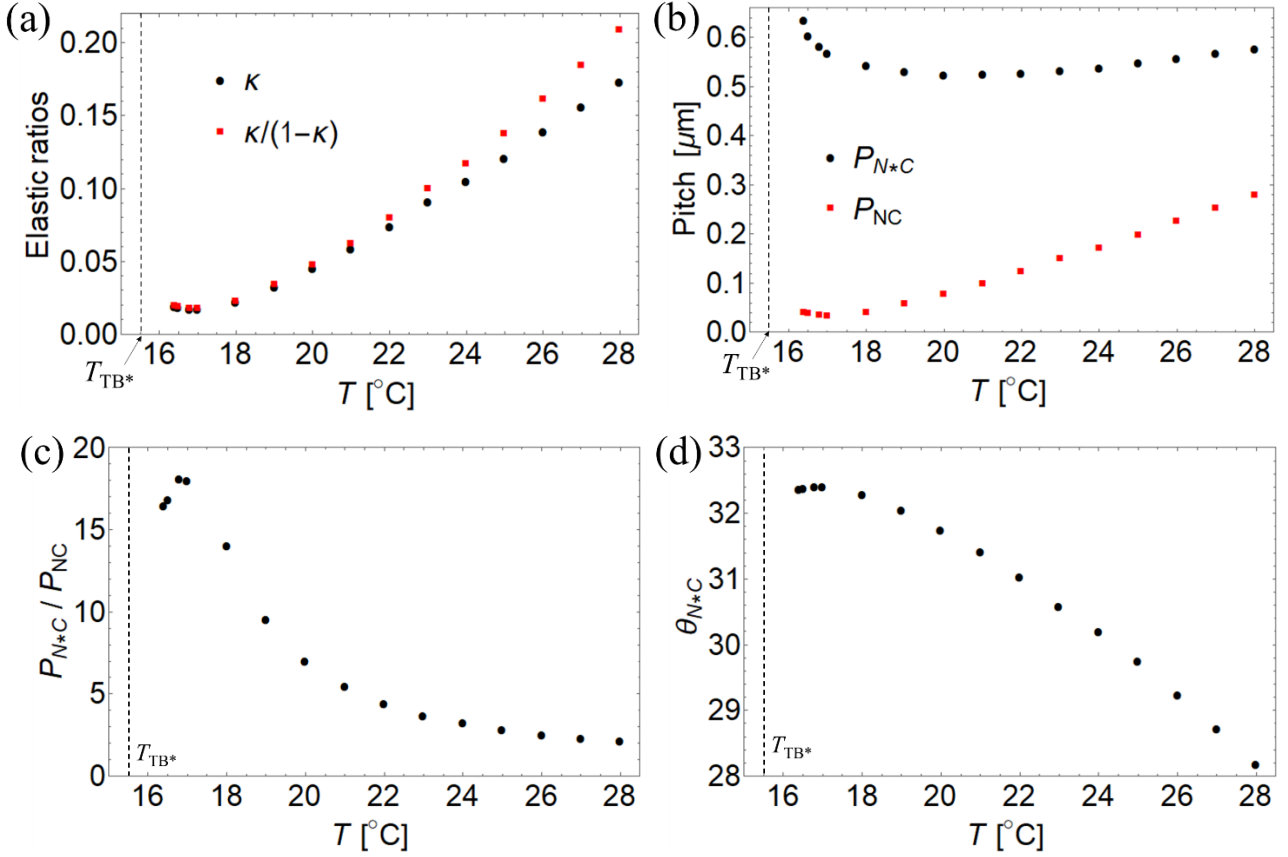

FIG.10. Temperature dependencies of (a) the elastic ratios $\kappa$ and $\kappa/(1-\kappa)$, (b) maximum $P_{N*C}$ and minimum $P_{NC}$ Ch$_{OH}$ pitches, (c) ratio of the pitches $P_{N*C}/P_{NC}$, and (d) maximum opening cone angle $\theta_{N*C}$.

## CONCLUSIONS

We measure the elastic constants of twist $K_{22}$ and bend $K_{33}$ directly in a chiral nematic, using electrically-tunable Bragg reflection of light from the Ch$_{OH}$ state and the critical field of unwinding the Ch$_{OH}$ structure, as well as the material properties such as dielectric anisotropy, off-field cholesteric pitch, and refractive indices. The measurements were made possible by introducing two independent techniques, based on capacitance and interference measurements, to determine the critical field $E_{NC}$, at which the Ch$_{OH}$ state unwinds into a uniform structure. The corresponding pitch at the fields approaching $E_{NC}$ is so small (decreasing below 50 nm) that seemingly straightforward approaches such as Bragg reflection in the visible part of the spectrum, or microscopy observations of textures, are not applicable. The proposed approach allowed us to demonstrate a very broad range of pitch tunability. For the studied mixture, we find that the ratio of the largest to smallest pitch in the electrically tunable range is about 18; a similar ratio is expected for the tunable Bragg reflection of light, from some $\lambda_{min}$ to $\lambda_{max} \approx 18\, \lambda_{min}$.

The ratio of elastic constants $\kappa = K_{33}/K_{22}$ is small, in the range 0.017-0.2 in a wide temperature range (>10°C) above the transition to the chiral twist-bend nematic phase. The smallness of $\kappa$ explains the broad temperature range of the existence of the $Ch_{OH}$ structure that allows one to continuously tune the wavelength of Bragg reflection by an external field. The temperature dependencies $K_{22}(T)$ and $K_{33}(T)$ allow us to characterize the $Ch_{OH}$ structure, namely, determine the range of the electrically tunable pitch $P$ and the maximum cone angle $\theta$. These dependencies are of interest on their own, as both $K_{22}$ and especially $K_{33}$ increase rapidly near the Ch-N*$_{TB}$ transition, Fig.9 (a). Similar behavior is known in nematics [17,23,27,28] and cholesterics [6,24] that show a transition to N$_{TB}$ or N*$_{TB}$. The increase of $K_{22}$ and $K_{33}$ is also well established in materials in which the nematic [29] or cholesteric [5] phase experiences a transition into a smectic A phase. Similar to the case of smectics, the nuclei of the N$_{TB}$ or N*$_{TB}$ phases make the bend and twist of the director difficult because these deformations are incompatible with the equidistance of the pseudolayered structure. Interestingly, $K_{22}$ shows a non-monotonous temperature dependence, with a minimum that precedes a rapid increase upon cooling, Fig.9 (a). Although unusual, this behavior is similar to the one observed for $K_{22}$ in the nematic phase of pure CB7CB when the temperature approaches the N$_{TB}$ phase [17]. Furthermore, the temperature dependence of the ratio $K_{33}/K_{22}$ in Fig.10(a) and in pure CB7CB [17] are qualitatively the same, showing that $K_{33}$ diverges more rapidly than $K_{22}$ near the transition to N$_{TB}$.

The intrinsic (field-free) pitch $P_0$ of the Ch increases as the temperature is lowered, Fig.5. Similar behavior of the Ch pitch is observed near the transition to the smectic A [5] and N*$_{TB}$ phase [24,30]. The trend could be again explained by the proximity to pseudo-layered N$_{TB}$ and N*$_{TB}$ phases, but Oswald and Dequidt [24] proposed an additional mechanism, rooted in the possibility that N*$_{TB}$ nuclei impose an opposite sense of chirality onto the surrounding Ch background [24]. Similarly to $K_{22}$, the linear chiral term $K_C = 2\pi\, K_{22}/P_0$ in the elastic energy density of Ch shows a nonmonotonic temperature behavior: $K_C$ is nearly constant far away from N*$_{TB}$, then shows a minimum and a rapid increase as the temperature is lowered towards N*$_{TB}$, Fig.9(b). The decrease might be qualitatively explained by the Oswald-Dequidt model [24], in which the Ch helicoid is partially unwound by N*$_{TB}$ nuclei (which implies an increase of $P_0$), followed by the strong increase of $K_{22}$.

## Acknowledgments

The work was supported by the NSF grant ECCS-1906104.